\documentclass[pra, a4paper, twocolumn, 10pt, showpacs]{revtex4}%, showpacs
\usepackage{bbm, amsmath, amssymb, amsthm, bm,textcomp, nicefrac,enumerate}%,mathrsfs
\usepackage[T1]{fontenc}
\usepackage[latin9]{inputenc}
\usepackage{graphicx}
\usepackage{geometry}

\begin{document}

\title{The kind of entanglement that speeds up quantum evolution}

\author{F.\ Fr\"owis}

\affiliation{Institut f\"ur Theoretische Physik, Universit\"at
  Innsbruck, Technikerstra\ss e 25, A-6020 Innsbruck,  Austria}
\date{\today}

\begin{abstract}
  The ``speed'' of unitary quantum evolution was recently shown to be connected to entanglement in multipartite quantum systems. Here, we discuss a tighter version of the Mandelstam-Tamm uncertainty relation that depends on the Fisher information. The passage time is estimated by a lower bound that depends inversely proportional
to the square root of the Fisher information. This leads to a better understanding of the origin of a fast quantum
time evolution of entangled states.
\end{abstract}


\maketitle

% ---------------------------------------------------
% Introduction 
% ---------------------------------------------------

\section{Introduction}
\label{sec:introduction}

The nontrivial aspects of time evolution in quantum mechanics have been fascinating and puzzling since the first days of its discovery. An important question is how quickly a quantum state evolves to a state that is distinguishable from the initial one with certainty. In the following, we refer to this concept as the ``speed of quantum evolution'' or ``quantum speed'' for short. It is expressed by the so-called survival probability wherein one uses the projection onto the initial quantum state for the discrimination. Quite general answers to this question have been found. These are not only interesting for purely academic reasons but are important contributions to practical issues like the possible speed of quantum gates in modern quantum-computation architectures. Already in 1945, Mandelstam and Tamm \cite{MandelstamTamm} showed that, for a pure state, the speed of quantum evolution is limited by the standard deviation of the system Hamiltonian. This bound was rediscovered via different reasoning by Fleming \cite{Fleming}. Given an initial pure state $\rho = \left| \psi \rangle\!\langle \psi\right|$ and a unitary time evolution under the time-independent Hamiltonian $H$, $\rho(t) = e^{-iHt/\hbar} \rho e^{iHt/\hbar}$, and the survival probability $P_{\rho}(t)\mathrel{\mathop:}=  \mathrm{Tr}\left[\rho \rho(t)  \right]$ was shown to be lower bounded by
\begin{equation}
\label{eq:1}
P_{\rho}(t) \geq \cos^2 \frac{\left( \Delta H\right)_{\rho} t}{\hbar} 
\end{equation}
during the time interval $\left( \Delta H\right)_{\rho}  \left| t \right|/\hbar \in [0,\pi/2]$. Here, $\left( \Delta H\right)_{\rho} = \sqrt{\mathrm{Tr}\left( H^2\rho \right)-\mathrm{Tr}\left( H\rho \right)^2}$ is the standard deviation of $H$.  The minimal time $\theta_{\bot}$ that passes before the initial state evolves into an orthogonal state is called the orthogonalization time or passage time. From Eq.~(\ref{eq:1}), it trivially follows that

\begin{equation}
\label{eq:2}
\theta_{\bot} \geq \frac{\pi\hbar}{2 \left( \Delta H\right)_{\rho}}.
\end{equation}

Recently, other general bounds and restrictions on the time evolution were established \cite{MargolusLevitin,Bounds}. For instance, it was shown by Margolus and Levitin \cite{MargolusLevitin} that the expectation value $\langle H \rangle_{\rho}= \mathrm{Tr}\left( H \rho \right)$ also gives a speed limit on the quantum evolution.

Furthermore, it was recognized \cite{Giovannetti03b} that entanglement can speed up time evolution in multipartite systems. If, for instance, the Hamiltonian is a sum of local terms, certain entangled states exhibit a passage time that is much shorter than any product state could have. This insight was followed by an intense study of the connection between the speed of evolution and entanglement \cite{ConnectionEntQuSpeed,EntSpeedMixed}.

A topic which has, on first sight, little in common with quantum speed is the theory of quantum metrology. The goal is to estimate an unknown parameter $\omega$ through measurements in an optimal way. The occurring error $\delta \omega$ of the estimate depends on how well we can distinguish two quantum states that differ in the actual value of $\omega$. This ``distinguishability'' is measured by the so-called Fisher information $\mathcal{F}$ \cite{ClassicalFisher,CramerRao,Helstrom,Holevo,BraunsteinCaves}, which gives rise to a lower bound on $\delta \omega$, called the Cram\'er-Rao bound \cite{ClassicalFisher,CramerRao,Helstrom,Holevo}. This bridges quantum metrology and quantum speed, if the parametrization is generated by the time evolution.

The relation between the two fields is accomplished by general insights into the geometric structure of quantum mechanics \cite{Wooters81,TimeEnergyUncertainty,GeometricMeaning}. This leads to a better understanding of the intrinsic structure of the Hilbert space and its relation to the physical interpretation. In addition, new results were established, e.g., the Margolus-Levitin bound \cite{MargolusLevitin} was used to prove a new inequality for parameter-estimation protocols \cite{Giovannetti11}.

In this paper, we reveal which kind of entanglement is useful to speed up the unitary evolution of generally mixed states $\rho$. To this end, we show that $(\Delta H)_{\rho}$ in Eqs.~(\ref{eq:1}) and (\ref{eq:2}) can be replaced by the Fisher information $\hbar \sqrt{\mathcal{F}(\rho,H)}/2 \leq (\Delta H)_{\rho}$. These findings give a clearer understanding of the role of entanglement for a potentially fast time evolution, since the connection of the Fisher information to entanglement is much more intimate than the relation between $(\Delta H)_{\rho}$ and the entanglement of $\rho$ \cite{EntSpeedMixed,pezze09,EntanglementFisher}. Note that the interplay among quantum speed, Fisher information and the geometrical interpretation of Hilbert space has been discussed by other authors \cite{Uhlmann,Holevo,BraunsteinCaves96}. However, explicit proofs and a clear discussion on the implications for entanglement were missing. 

For the remainder of this section, we set the framework to improve the Mandelstam-Tamm relations (\ref{eq:1}) and (\ref{eq:2}). In particular, we introduce different generalizations of $P_{\rho}(t)$ for mixed states and review the theory of quantum metrology. In Sec.~\ref{sec:deriv-impr-mand}, we give the proof of the new bound. We discuss the consequences for the role of entanglement for time evolution in Sec.~\ref{sec:conn-entangl}. In this section, we also extend a particular form of the survival probability to more general projective measurements.

\subsection{Survival probability for mixed states}
\label{sec:surv-prob-mixed}

Throughout this paper, we exclusively consider finite-dimensional Hilbert spaces $\mathcal{H}$. Furthermore, we restrict ourselves to time-independent Hamiltonians $H$. Given a time-evolved state $\left| \psi(t) \right\rangle  = e^{-iHt/\hbar} \left| \psi \right\rangle \in \mathcal{H}$, the transition probability $P_{\rho}(t)=\left| \langle \psi | \psi(t)\rangle  \right|^2 \in \left[ 0,1 \right]$ is a well-accepted measure to quantify the speed with which a quantum state evolves away from the initial state vector. The generalization to mixed states is not unique. At least two possibilities have been discussed in the literature. On the one hand \cite{Giovannetti03a,Uhlmann}, one can use the square of the fidelity between the initial density operator $\rho$ and the time evolved $\rho(t)= e^{-iHt/\hbar} \rho e^{iHt/\hbar}$, as
\begin{equation}
T_{\rho}(t)\mathrel{\mathop:}=\left(  \mathrm{Tr}\sqrt{\sqrt{\rho}\rho(t) \sqrt{\rho}}\right)^2\label{eq:9}.
\end{equation}
For generic mixed states, Eq.~(\ref{eq:9}) can not be expressed as an expectation value of a single observable. In contrast, the orthogonal projection $\Pi_{\rho}$ onto the range of $\rho$ is a realizable observable. The definition for this version of the survival probability 
\begin{equation}
E_{\rho}(t)\mathrel{\mathop:}=\mathrm{Tr}\left[ \Pi_{\rho}\rho(t)\right]\label{eq:10}
\end{equation}
was proposed in Ref.~\cite{FrGrPe08}. If $\rho$ is of full rank, $E_{\rho}(t) = 1$ for all times because $\Pi_{\rho} = \mathbbm{1}$ is the identity. To circumvent this drawback, we can choose an observable $A$ that maximizes the difference $\left| \langle A \rangle_{\rho(t)}-\langle A \rangle_{\rho} \right|$. This leads to the trace distance $d(\rho,\sigma) =  \lVert \rho-\sigma \rVert_1/2$ \cite{TraceNorm} since $\max_A \mathrm{Tr}\left[ A \left( \rho-\sigma \right) \right] = d(\rho,\sigma)$ if the maximum is taken over all observables with spectral radius 1/2 \cite[p. 404]{NielsenChuang}. The corresponding generalization of the survival probability reads
\begin{equation}
D_{\rho}(t) \mathrel{\mathop:}= 1 - \frac{1}{4} \lVert \rho(t)-\rho \rVert^2_1.\label{eq:11}
\end{equation}
Note that $T_{\rho}(t) \leq D_{\rho}(t)$ \cite[p. 415]{NielsenChuang}. For this reason, lower bounds on $T_{\rho}(t)$ directly apply to $D_{\rho}(t)$.

In the following, we call all three definitions (\ref{eq:9}), (\ref{eq:10}) and (\ref{eq:11}) survival probabilities. For a pure state $\rho=\left|\psi  \rangle\!\langle \psi\right| $, they coincide with $P_{\rho}(t)$. The obvious generalization of Eq.~(\ref{eq:1}) for $T_{\rho}(t)$ and $E_{\rho}(t)$ was proved in Refs.~\cite{Giovannetti03a,Uhlmann} and \cite{FrGrPe08}, respectively. All three approaches, therefore, have the common lower bound (\ref{eq:2}).

For general $\rho$, this bound is not very tight. An extreme case is an incoherent sum of eigenstates of $H$. Then $T_{\rho}(t)=E_{\rho}(t) = 1$ for all times, whereas the bound (\ref{eq:2}) may suggest a rather quick evolution. This observation was discussed in Ref.~\cite{EntSpeedMixed} and motivates an improvement of Eqs.~(\ref{eq:1}) and (\ref{eq:2}) in the case of mixed states.

\subsection{Quantum Fisher information}
\label{sec:quant-fish-inform}

The last part of the introduction reviews the properties of the Fisher information in relation to the problem of distinguishing density operators. Suppose we want to estimate an unknown parameter $\omega$ that is encoded in a system we have access to. To reveal its value, we perform a measurement with possible outcomes $a_i$, which here are supposed to be discrete. The relative frequencies of the outcomes depend on $\omega$ and are associated with a probability distribution $p_i(\omega)$ on a probability space that is given through our choice of measurement. From $p_i(\omega)$ we try to estimate $\omega$ as accurately as possible. The error $\delta\omega$ we have to assume for a certain estimator is not simple to calculate, but, for an unbiased estimate, can be bounded from below by the Cram\'er-Rao bound \cite{ClassicalFisher,CramerRao} $\delta\omega \geq 1/\sqrt{nF(\omega)}$, where $n$ is the number of repetitions of the experiment and
\begin{equation}
\label{eq:3}
F(\omega) = \sum_i p_i(\omega) \left[ \frac{d}{d\omega}\ln p_{i}(\omega) \right]^2
\end{equation}
is the Fisher information, here also referred to as classical Fisher information. 

So far, the formalism of quantum mechanics was not used explicitly. We now assume that a density operator $\rho$ represents the system under consideration. Hence, $\rho$ depends on $\omega$. The probabilities are calculated from the expectation values $p_i(\omega) = \mathrm{Tr}[E_i\rho(\omega)]$ of a chosen measurement $\left\{ E_i \right\}_i$. If we vary our measurement to minimize the estimated error $\delta\omega$, we end up with the so-called quantum Fisher information $\mathcal{F}(\omega) = \max_{ \left\{ E_i \right\}}F(\omega)$. In addition to its importance in the theory of quantum metrology, the quantum Fisher information also gives insight into the structure of the space of density operators, as highlighted by Braunstein and Caves \cite{BraunsteinCaves}. The different values of $\omega$ parametrize a curve through this space. One can define a ``distinguishability'' metric via $ds = \frac{1}{2}\sqrt{\mathcal{F}} d\omega$. This was shown \cite{Uhlmann,BraunsteinCaves} to be equivalent to the Bures distance \cite{Bures} between the two states $\rho(\omega)$ and $\rho(\omega+\delta\omega)$.

In this context, the parametrization through the space of density operators is generated by a time-independent Hamiltonian. The parameter is the elapsing time. Then one can show \cite{Helstrom,BraunsteinCaves} that the Fisher information is time independent and it has the form
\begin{equation}
\label{eq:7}
\mathcal{F}(\rho,H) = 2\sum_{i,j} \frac{\left( \pi_i-\pi_j \right)^2}{\pi_i+\pi_j}\left| \left\langle i \right| H/ \hbar \left| j \right\rangle  \right|^2,
\end{equation}
where we used the spectral decomposition of the initial density operator 
\begin{equation}
\rho = \sum_i \pi_i \left| i \rangle\!\langle i \right|.\label{eq:18}
\end{equation}
Note that $\mathcal{F}(\rho,H)$ can be further estimated as $\mathcal{F}(\rho,H)\leq 4 \left( \Delta H \right)^2_{\rho}/\hbar^2$ \cite{Uhlmann,BraunsteinCaves}. The equality holds if $\rho$ is pure.

\section{Improved Mandelstam-Tamm bound}
\label{sec:deriv-impr-mand}

In this section, we prove a version of the Mandelstam-Tamm inequality in which $\left( \Delta H \right)_{\rho}$ in Eqs.~(\ref{eq:1}) and (\ref{eq:2}) is replaced by $\hbar \sqrt{\mathcal{F}(\rho,H)}/2$. The new relations are identical to the original relations in the case of pure states but are generically tighter for mixed states. We present proofs for both types of survival probabilities $T_{\rho}(t)$ for Eq.~(\ref{eq:9}) and for Eq.~$E_{\rho}(t)$ (\ref{eq:10}) in the following paragraphs.

\subsection{Bound on $T_{\rho}(t)$ and $D_{\rho}(t)$}
\label{sec:proof-t_rhot}

We prove the following proposition. For the time interval $0\leq\sqrt{\mathcal{F}(\rho,H)}\left| t \right| \leq \pi$, the transition probability $T_{\rho}(t)$ can be bounded from below by
\begin{equation}
  \label{eq:12}
  T_{\rho}(t) \geq \cos^2  \sqrt{\mathcal{F}(\rho,H)}t/2.
\end{equation}
The proof of Eq.~(\ref{eq:12}) can implicitly be found in Ref.~\cite{Uhlmann} and, due to its elegant simplicity, is sketched here. The two density operators $\rho$ and $\rho(t)$ lie on a curve generated by the Hamiltonian $H$. The length of this path $\gamma$ between these states is $S=\int_{\gamma} ds$. The metric we take here is the Bures metric \cite{Bures}. We use $ds = \frac{1}{2}\sqrt{\mathcal{F}(\rho,H)} dt$ and the time independence of $\mathcal{F}(\rho,H)$ to find $S = \int_0^t \frac{1}{2}\sqrt{\mathcal{F}(\rho,H)} dt^{\prime} = \frac{1}{2}\sqrt{\mathcal{F}(\rho,H)} t$. On the other hand, we can express the distance $S_0 \leq S$ between two density operators in terms of the angle $\arccos\sqrt{T_{\rho}(t)}$ \cite{Uhlmann86}. This follows if we consider the purifications $\left| \phi \right\rangle $ and $\left| \phi(t) \right\rangle $ of $\rho$ and $\rho(t)$, respectively, that give the maximal overlap $\left| \langle \phi | \phi(t)\rangle  \right|$. The distance between them is the geodesic arc connecting $\left| \phi \right\rangle $ and $\left| \phi(t) \right\rangle $ on the unit sphere, whose length is $S_0=\arccos\left| \langle \phi | \phi(t)\rangle \right| = \arccos\sqrt{T_{\rho}(t)}$ \cite{Uhlmann}. So we find that $S\geq \arccos\sqrt{T_{\rho}(t)}$. For the time interval $0\leq\sqrt{\mathcal{F}(\rho,H)}\left| t \right| \leq \pi$, we can invert the relation and find Eq.~(\ref{eq:12}).

Since $D_{\rho}(t) \geq  T_{\rho}(t)$, we have for the same time interval $0\leq\sqrt{\mathcal{F}(\rho,H)}\left| t \right| \leq \pi$ that
\begin{equation}
  \label{eq:12a}
  D_{\rho}(t) \geq \cos^2  \sqrt{\mathcal{F}(\rho,H)}t/2.
\end{equation}
\subsection{Bound on $E_{\rho}(t)$}
\label{sec:proof-t_rhot}

We now derive a differential inequality that allows us to prove a tighter version of the Mandelstam-Tamm inequality for $E_{\rho}(t)$. The system $\rho(t)$ is measured by means of the orthogonal projection $\Pi_{\rho}$ onto the range of $\rho\equiv \rho(0)$. We calculate the classical Fisher information of Eq.~(\ref{eq:3}). The regarded ``parameter'' is the time itself, $\omega \equiv t$. With the two outcomes $p_1 \equiv p = \mathrm{Tr}[\Pi_{\rho} \rho(t)] = E_{\rho}(t)$ and $p_2 = 1 - p$, Eq.~(\ref{eq:3}) has the simple form
\begin{equation}
F(t) = \frac{[\dot{p}(t)]^2} { p(t) \left[ 1-p(t) \right] }.\label{eq:20}
\end{equation}
Noting that with $p(t) \left[ 1-p(t) \right] = \left( \Delta \Pi_{\rho} \right)_{\rho(t)}^2$, one has $\left| \dot{p}(t) \right| = \sqrt{F(t)} \left( \Delta \Pi_{\rho} \right)_{\rho(t)}$. Next, the classical Fisher information is estimated from above by the quantum version $\mathcal{F}$, which is time-independent here. We therefore have
\begin{equation}
\label{eq:4}
\left| \frac{d}{dt}E_{\rho}(t) \right| \leq \sqrt{\mathcal{F}(\rho,H)}\left( \Delta \Pi_{\rho} \right)_{\rho(t)}.
\end{equation}
Equation (\ref{eq:4}) states that the maximal rate with which $E_{\rho}(t)$ can change is limited by the quantum Fisher information. The attainability of this bound depends strongly on the observable $\Pi_{\rho}$. The more suitable it is to distinguish two neighboring states $\rho(t)$ and $\rho(t+d t)$, the tighter is bound Eq.~(\ref{eq:4}).

The differential inequality (\ref{eq:4}) has to be contrasted to the original inequality that leads to the Mandelstam-Tamm relation (\ref{eq:1}). We directly calculate the time derivative of $E_{\rho}(t)$ to get $\left|\frac{d}{dt} E_{\rho}(t) \right| = \left| \mathrm{Tr}\left( \left[\rho(t),H/\hbar  \right]\Pi_{\rho} \right) \right| = 1/\hbar \left|\langle \left[ H,\Pi_{\rho} \right] \rangle_{\rho(t)}  \right| $. With the help of the Heisenberg uncertainty relation \cite{Heisenberg}, we find
\begin{equation}
\label{eq:5}
\left|\frac{d}{dt} E_{\rho}(t) \right| \leq \frac{2}{\hbar} \left( \Delta H \right)_{\rho}\left( \Delta \Pi_{\rho} \right)_{\rho(t)}.
\end{equation}
Recalling that $\mathcal{F}(\rho,H)\leq 4 \left( \Delta H \right)^2_{\rho}/\hbar^2$, one can see that the general inequality (\ref{eq:4}) constitutes an improvement of Eq.~(\ref{eq:5}) whereas they are identical for pure states.

From Eq.~(\ref{eq:4}) we now deduce a general lower bound on $E_{\rho}(t)$. Since both $\left( \Delta H \right)_{\rho}$ and $\mathcal{F}(\rho,H)$ are time independent, Eq.~(\ref{eq:4}) can be treated exactly in the same way as Eq.~(\ref{eq:5}); for a detailed discussion, see Ref.~\cite{FrGrPe08}. There, a slightly sharper version of Eq.~(\ref{eq:1}) was found. We formulate it here for the Fisher information. It states that for the survival probability $E_{\rho}(t)$, only one of the two cases:
\begin{enumerate}[(i)]\item \label{item:1}
  \begin{subequations}
    \label{eq:6}
     $0<\sqrt{\mathcal{F}(\rho,H)}\left| t \right| \leq \pi:$
    \begin{equation}\label{eq:6a}
      E_{\rho}(t) > \cos^2 \sqrt{\mathcal{F}(\rho,H)}t/2
    \end{equation}
    \item $\forall t \in \mathbbm{R}:$
    \begin{equation}
      E_{\rho}(t) =\cos^2 \sqrt{\mathcal{F}(\rho,H)}t/2
    \end{equation}
  \end{subequations}
\end{enumerate}

is realized. This means that if, for a time $t>0$ we find situation (\ref{item:1}) $E_{\rho}(t)$ cannot come back to the bound $\cos^2 \sqrt{\mathcal{F}(\rho,H)}t/2$ because such a drastic change of $E_{\rho}(t)$ is forbidden by Eq.~(\ref{eq:4}).

\section{Discussion}
\label{sec:conn-entangl}

We see that independent of the specific form for the survival probability we choose [\textit{cf.} Eqs.~(\ref{eq:9}), (\ref{eq:10}) and (\ref{eq:11})], the same lower bound can be derived for all three versions. Hence, it directly follows that the orthogonalization time for all three definitions can be bounded from below by
\begin{equation}
\label{eq:8}
\theta_{\bot} \geq \frac{\pi }{\sqrt{\mathcal{F}(\rho,H)}}.
\end{equation}
This improved Mandelstam-Tamm uncertainty relation is very similar to the aforementioned Cram\'er-Rao bound for the error $\delta\omega$ in the parameter estimation theory. Although the bounds differ in the specific meaning, their similarity alludes to the relation between quantum metrology and survival probability on the common ground of distinguishing density operators.

We shortly address the question of the attainability of Eq.~(\ref{eq:8}). For pure states, it has been shown (e.g., by Refs.~\cite{GeometricMeaning,Giovannetti03a,FrGrPe08}) that equality holds if and only if the initial state is an equally weighted superposition of exactly two eigenstates of $H$. For mixed states, a similar result can be proved. As in Ref.~\cite{FrGrPe08}, we explicitly write out $E_{\rho}(t)$ with respect to the spectral decomposition of $\rho$ of Eq.~(\ref{eq:18}). We find
\begin{equation}
\label{eq:19}
E_{\rho}(t) = \sum_i\pi_i P_{\left| i \rangle\!\langle i\right| }(t) + \sum_{i \neq j }\pi_i \left| \left\langle j \right| e^{-iHt/\hbar}\left| i \right\rangle  \right|^2.
\end{equation}
Equality in Eqs.~(\ref{eq:2}) and (\ref{eq:8}) hold if all terms in the second sum of Eq.~(\ref{eq:19}) vanish and, for all $i$, $P_{\left| i \rangle\!\langle i\right|} (\theta_{\bot}) =0$ is simultaneously fulfilled. In these cases,  $\mathcal{F}(\rho,H)= 4 \left( \Delta H \right)^2_{\rho}/\hbar^2$, which is necessary, because Eq.~(\ref{eq:2}) cannot be improved further. So far, there are no examples known in which there is equality in Eq.~(\ref{eq:8}) but not in Eq.~(\ref{eq:2}), except when $\rho$ commutes with $H$. Then $\mathcal{F}(\rho,H) = 0$, meaning that there is no finite passage time.

\subsection{Entanglement and quantum speed}
\label{sec:entangl-quant-speed}

We now discuss the importance of entanglement for the speed of quantum evolution. Let us consider multipartite-qubit states on the Hilbert space $\mathcal{H} = \mathbbm{C}^{2\otimes N}$, where $N$ is the number of qubits. The Hamiltonian consists of a sum of one-particle terms,
for which every term exhibits a constant operator norm $\epsilon$. While for general states there is no direct connection between the entanglement between the particles and $\left( \Delta H \right)_{\rho}$, the relation between entanglement and quantum Fisher information is more stringent. One can show \cite{pezze09} that if $\mathcal{F}(\rho,H) > 4\epsilon^2 N/\hbar^2$, then $\rho$ is entangled and is, in principle, more useful for parameter estimation (in the sense of Sec.~\ref{sec:quant-fish-inform}) than any nonentangled state. Very recently, in Ref.~\cite{EntanglementFisher} a connection between $\mathcal{F}$ and the so-called $k$-producible states has been established. Let us divide $\mathcal{H}$ into subsets of at most $k\leq N$ qubits, that is, $\mathcal{H} = \mathbbm{C}^{2\otimes k_1} \otimes \ldots \otimes \mathbbm{C}^{2\otimes k_m}$ with $k_i \leq k$ and $\sum_ik_i = N$. A pure state $\left| \psi ^{(k)}\right\rangle  = \left| \psi_1 \right\rangle \otimes \ldots \otimes \left| \psi_m \right\rangle $ with $\left| \psi_i \right\rangle \in \mathbbm{C}^{2\otimes k_i}$ is called $k$-producible. A general state $\rho^{(k)}$ is $k$-producible if it is a mixture of $k$-producible pure states with respect to different partitions $\left\{k_i  \right\}_i$. The authors of Ref.~\cite{EntanglementFisher} found that --for any $H$ as a sum of one-particle terms-- $\mathcal{F}(\rho^{(k)},H/(2\epsilon)) \leq s k^2 + \left( N-sk \right)^2$ with $s = \lfloor \frac{N}{k} \rfloor$. This means that the degree of entanglement with respect to this classification gives a direct upper bound on the speed of evolution
\begin{equation}
\label{eq:13}
\theta_{\bot} \geq \frac{\pi \hbar}{2\epsilon\sqrt{s k^2 + \left( N-sk \right)^2}}.
\end{equation}
Note that the bound on the orthogonalization time is now independent of the specific local $H$. These results imply that entanglement measured by the Fisher information plays an essential role for a potentially fast time evolution.

\subsection{Explicit example}
\label{sec:explicit-example}

\begin{figure}[tb] 
    \includegraphics[width=\columnwidth]{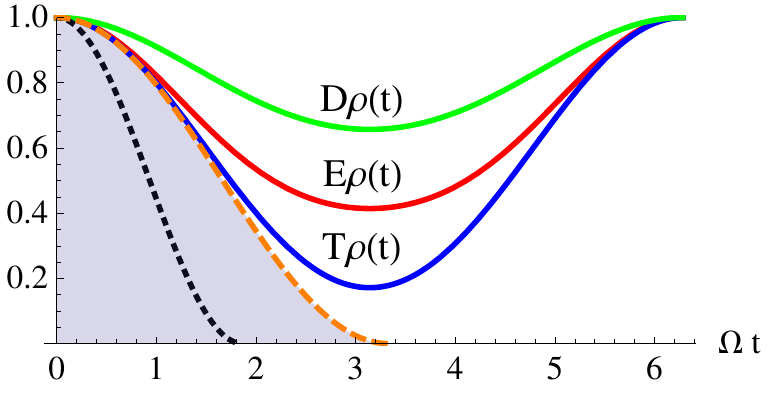}
\caption[]{\label{fig:2qubits} (Color online) The two-qubit example that is discussed in Sec.~\ref{sec:explicit-example} is plotted for $x=\sqrt{2}-1$. The solid curves from top to bottom are $D_{\rho}(t)$, $E_{\rho}(t)$ and $T_{\rho}(t)$. The black, dotted curve shows $\cos^2 \left( \Delta H \right)_{\rho}t/\hbar$ from the Mandelstam-Tam inequality (\ref{eq:1}). The orange, dashed curve above the shading represents the improved version $\cos^2 \sqrt{\mathcal{F}(\rho,H)} t/2$ from Eqs.~(\ref{eq:12}), (\ref{eq:12a}) and (\ref{eq:6}), which are clearly tighter than Eq.~(\ref{eq:1}).}\end{figure}

We examine Eqs.~(\ref{eq:12}), (\ref{eq:12a}) and (\ref{eq:6}) for two qubits, i.e., the Hilbert space is $\mathbbm{C}^4$. We start with the parametrized state $\rho = (1-x) \left| 00 \rangle\!\langle 00\right| + x \left| \psi^{+} \rangle\!\langle \psi^{+}\right| $, which was already discussed in Ref.~\cite{EntSpeedMixed}. The two basis states $\left| 0 \right\rangle $ and $\left| 1 \right\rangle \in \mathbbm{C}^2$ are the eigenstates of the Pauli $\sigma_z$ operator and $\left| \psi^{+} \right\rangle  = \frac{1}{\sqrt{2}}\left( \left| 01 \right\rangle + \left| 10 \right\rangle  \right)$ is a maximally entangled state. We use the parameter $x \in \left[ 0,1 \right]$ to mix $\left| 00 \right\rangle $ and $\left| \psi^{+} \right\rangle $. The dynamics are generated by the Hamiltonian $H = \frac{1}{2}\hbar \Omega \left( \sigma_x^{(1)} + \sigma_x^{(2)} \right)$, $\sigma_x = \left| 0 \rangle\!\langle 1\right| + \left| 1 \rangle\!\langle 0\right| $. The standard deviation reads $\left( \Delta H \right)_{\rho} = \hbar \Omega \sqrt{(1+x)/2}$, while for the Fisher information (\ref{eq:7}) we find $\mathcal{F}(\rho,H) = 2 \Omega^2 \left( 1-3x +4x^2 \right)$. We see that $\mathcal{F}(\rho,H)$ coincides with $4\left( \Delta H \right)_{\rho}^2/\hbar^2$ for $x=0,1$, as it should. For all other $x$, $\mathcal{F}(\rho,H)$ is strictly smaller than $4\left( \Delta H \right)_{\rho}^2/\hbar^2$. The maximal difference is attained for $x=\sqrt{2}-1$. For this value, we calculate $E_{\rho}(t)$, $F_{\rho}(t)$ and $D_{\rho}(t)$. Instead of giving the lengthy analytical expressions, we show the plot in Fig.~\ref{fig:2qubits} for a certain time interval. There, we compare the quantities with the Mandelstam-Tamm bound (\ref{eq:1}) with Eqs.~(\ref{eq:12}), (\ref{eq:12a}) and (\ref{eq:6}). We see that the improved bounds are much tighter than those of Eq.~(\ref{eq:1}).

\subsection{General projective measurement}
\label{sec:disc-furth-gener}

We now consider other projective measurements than $\Pi_\rho$ as observables for $E_{\rho}(t)$. This is motivated by the observation that for initial states with high rank, the projection onto a high-dimensional subspace often leads to survival probabilities staying close to one. If we encounter a spectrum of $\rho$ with some eigenvalues close to zero, we may use a projective measurement that ignores the space spanned by the corresponding eigenvectors.  The derivation of Eq.~(\ref{eq:4}) does not rely on a certain structure of the measurement, as long as it is an orthogonal projection, which we simply denote by $\Pi$. What changes in our reasoning is that we do have different initial conditions $\langle \Pi  \rangle_{\rho(0)} = c \in \left[ 0,1 \right]$. One can easily verify that the differential inequality
\begin{equation}
\frac{d}{dt}E_{\rho}(t) \geq - \sqrt{\mathcal{F}(\rho,H)}\left( \Delta \Pi \right)_{\rho(t)}\label{eq:15}
\end{equation}
leads to
\begin{equation}
E_{\rho}(t) \geq \cos^2 \left(\sqrt{\mathcal{F}(\rho,H)}t/2 +\delta  \right)\label{eq:14}
\end{equation}
for all $t \in  \left[ 0,(\pi-2\delta)/\sqrt{\mathcal{F}(\rho,H)} \right]$ with $\delta = \arccos \sqrt{c}$. Similarly, the inequality
\begin{equation}
\frac{d}{dt}E_{\rho}(t) \leq \sqrt{\mathcal{F}(\rho,H)}\left( \Delta \Pi \right)_{\rho(t)}\label{eq:16}
\end{equation}
gives
\begin{equation}
E_{\rho}(t) \leq \sin^2 \left(\sqrt{\mathcal{F}(\rho,H)}t/2 +\delta^{\prime}  \right)\label{eq:17}
\end{equation}
for all $t \in  \left[ 0,(\pi-2\delta^{\prime})/\sqrt{\mathcal{F}(\rho,H)} \right]$ with $\delta^{\prime} = \arcsin \sqrt{c}$.

\begin{figure}[tb]
  \begin{picture}(210,105)
    \put(110,-5){\includegraphics[width=120pt]{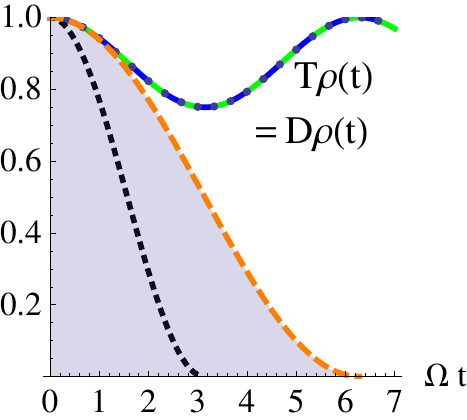}}
    \put(-15,-8){\includegraphics[width=120pt]{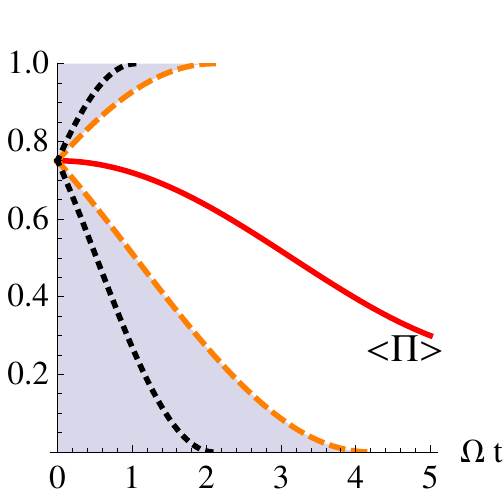}}
    \put(40,103){(a)}
\put(158,103){(b)}
\end{picture}
\caption[]{\label{fig:1qubit} (Color online) Illustration of the example in Sec.~\ref{sec:disc-furth-gener}. (a) The expectation value $\langle \Pi \rangle_{\rho(t)}$ with $\Pi=\left| 0 \rangle\!\langle 0\right| $ is plotted for $x=3/4$. The dashed, orange curves are the generalized lower and upper bounds from Eqs.~(\ref{eq:14}) and (\ref{eq:17}), respectively. These bounds can be compared to similar bounds (black, dotted curve) on the basis of Eq.~(\ref{eq:5}). (b) The same example for the survival probabilities $T_{\rho}(t)$ and $D_{\rho}(t)$, which coincide in this example. Again, we compare them to $\cos^2 \left( \Delta H \right)_{\rho}t/\hbar$ from Eq.~(\ref{eq:1}) (black, dotted curve) and  $\cos^2 \sqrt{\mathcal{F}(\rho,H)} t/2$ from Eq.~(\ref{eq:6}) (orange, dashed curve). In contrast to (a), the latter bound is tight for a small time interval.}
\end{figure}

We discuss an example for the generalized bounds (\ref{eq:14}) and (\ref{eq:17}). Suppose we have a one-qubit system that is a mixture of $\left| 0 \right\rangle $ and $\left| 1 \right\rangle $, $\rho = (1-x)\left| 0 \rangle\!\langle 0\right|  + x \left| 1 \rangle\!\langle 1\right| $, again with $x \in \left[ 0,1 \right]$. The Hamiltonian is chosen to be $H=\hbar \Omega\sigma_x /2 $. Any $x$ within the interval $\left( 0,1 \right)$ results in $\Pi_{\rho} = \mathbbm{1}$. To see a nontrivial time evolution, we choose $\Pi = \left| 0 \rangle\!\langle 0\right| $. The time dependent quantity $\langle \Pi \rangle_{\rho(t)}=\mathrm{Tr}\left[ \Pi \rho(t) \right]$ is plotted for $x=3/4$ in Fig.~\ref{fig:1qubit} (a). We see that due to the different time derivatives of $\langle \Pi \rangle_{\rho(t)}$ and its bounds at $t=0$, the bounds are not very tight. For the sake of completeness, we plot $T_{\rho}(t)$ and $D_{\rho}(t)$ for the same value $x$ in Fig.~\ref{fig:1qubit} (b). They coincide in this example.

\section{Conclusion}
\label{sec:conclusion-outlook}

In summary, we have discussed an improved Mandelstam-Tamm inequality [\textit{cf.}~(\ref{eq:12}), (\ref{eq:12a}) and (\ref{eq:6})] for different notions of survival probabilities, giving a common upper bound on the speed with which a quantum state can evolve. The new bound is in general tighter than the original bound (\ref{eq:1}) for mixed states. The Fisher information replaces the variance of the energy, also meaning that the new bound is more difficult to calculate. In return, we now understand better the role of entanglement in the context of ``quantum speed''. Not every kind of entanglement is useful to accelerate the time evolution compared to non-entangled states. It is necessary to exhibit a high Fisher information to speed up the time evolution compared to non-entangled states. With the upper bound \cite{EntanglementFisher} on $\mathcal{F}$ for $k$-producible quantum states, we can directly connect the survival probability with $k$.

In addition, the findings highlight the interplay of time evolution, quantum metrology and the geometry of quantum mechanics. The orthogonalization (or passage) time is lower bounded by the inverse of $\sqrt{\mathcal{F}(\rho,H)}$. The form of this uncertainty relation is very similar to the famous Cram\'er-Rao bound. Furthermore, a different kind of uncertainty relation has appeared in this context. Comparing Eq.~(\ref{eq:4}) with the derivation of Eq.~(\ref{eq:5}) gives $ \left| \langle \left[ H,\Pi \right] \rangle_{\rho(t)} \right| \leq \hbar\sqrt{\mathcal{F}(\rho,H)} \left( \Delta\Pi \right)_{\rho(t)}$. This is tighter than the Heisenberg uncertainty $\left| \langle \left[ H,\Pi \right] \rangle_{\rho(t)} \right| \leq 2 \left( \Delta H \right)_{\rho} \left( \Delta\Pi \right)_{\rho(t)}$. 

Finally, the study has been extended to arbitrary projective measurements for $E_{\rho}(t)$ and we have seen that a different choice of measurement can potentially reveal the dynamics of $\rho(t)$, if the projection onto the range of the initial state results in a trivial time evolution.

An open question is the attainability of the bound (\ref{eq:8}) for situations where $\mathcal{F}(\rho,H) < 4 \left( \Delta H \right)_{\rho}^2/\hbar^2$.

\section*{Acknowledgements}
\label{sec:acknowledgements}

The author is grateful for discussions with Lorenzo Maccone, Gebhard Gr\"ubl and Wolfgang D\"ur. The research was funded by the Austrian Science Fund (FWF):  P20748-N16, P24273-N16, SFB F40-FoQus F4012-N16 and the European Union (NAMEQUAM).
% ---------------------------------------------------
% ---------------------------------------------------


\begin{thebibliography}{99}

  %%%%% Introduction %%%%%

\bibitem[]{MandelstamTamm}L. Mandelstam and I. Tamm, J. Phys. (USSR) \textbf{9}, 249 (1945). 

\bibitem[]{Fleming}G. N. Fleming, %A Unitarity Bound on the Evolution of Nonstationary States
Nuovo Cimento A \textbf{16}, 232-240 (1973).

\bibitem[]{MargolusLevitin}N. Margolus and L. B. Levitin, %The Maximum Speed of Dynamical Evolution
Phys. D (Amsterdam, Neth.) \textbf{120}, 188-195 (1998).

\bibitem[]{Bounds}M. Andrews, %Bounds to Unitary Evolution
Phys. Rev. A \textbf{75}, 062112 (2007);
H. F. Chau, %Tight Upper Bound of the Maximum Speed of Evolution of a Quantum State
Phys. Rev. A \textbf{81}, 062133 (2010).



\bibitem[]{Giovannetti03b}V. Giovannetti, S. Lloyd, and L. Maccone, %The Role of Entanglement in Dynamical Evolution
Europhys. Lett. \textbf{62}, 615-621 (2003).

\bibitem[]{ConnectionEntQuSpeed}J. Batle , M. Casas, A. Plastino, and A. R. Plastino, %Connection Between Entanglement and the Speed of Quantum Evolution
  Phys. Rev. A \textbf{72}, 032337 (2005);
  A. Borr\'as, M. Casas, A. R. Plastino, and A. Plastino, %Entanglement and the Lower Bounds on the Speed of Quantum Evolution
Phys. Rev. A \textbf{74}, 022326 (2006);
S. Curilef, C. Zander, and A. R. Plastino, %Two Particles in a Double Well: Illustrating the Connection Between Entanglement and the Speed of Quantum Evolution
European Journal of Physics \textbf{27}, 1193-1203 (2006);
C. Zander, A. R. Plastino, A. Plastino, and M. Casas, %Entanglement and the Speed of Evolution of Multi-partite Quantum Systems
J. Phys. A: Math. Theor. \textbf{40}, 2861-2872 (2007);
A. Borras, A. R. Plastino, J. Batle, C. Zander, M. Casas, and A. Plastino, %Multiqubit Systems: Highly Entangled States and Entanglement Distribution
Journal of Physics A: Mathematical and Theoretical \textbf{40}, 13407-13421 (2007);
S. Curilef, C. Zander, and A. R. Plastino, %Speed of Quantum Evolution of Entangled Two Qubits States: Local Vs. Global Evolution
Journal of Physics: Conference Series \textbf{134}, 012003 (2008).

\bibitem[]{EntSpeedMixed}J. Kupferman and B. Reznik, %Entanglement and the Speed of Evolution in Mixed States
Phys. Rev. A \textbf{78}, 042305 (2008).

\bibitem[]{ClassicalFisher}R. A. Fisher, Math. Proc. Cambridge Philos. Soc. \textbf{22}, 700 (1925).
  
\bibitem[]{CramerRao} H. Cram\'er, \textit{Mathematical Methods of Statistics} (Princeton University Press, Princeton, 1946), p. 500.

\bibitem[]{Helstrom}C. W. Helstrom, \textit{Quantum Detection and Estimation Theory} (Academic, New York, 1976).

\bibitem[]{Holevo}
A. S. Holevo, \textit{Probabilistic and Statistical Aspects of Quantum Theory} (North-Holland Publishing, Amsterdam, 1982).

\bibitem{BraunsteinCaves}S.~L.~Braunstein and C.~M.~Caves, Phys. Rev. Lett. \textbf{72}, 3439 (1994).

  \bibitem{Wooters81} W.~K.~Wootters, Phys. Rev. D \textbf{23}, 357 (1981).

\bibitem[]{TimeEnergyUncertainty}J. Anandan and Y. Aharonov, %Geometry of Quantum Evolution
Phys. Rev. Lett. \textbf{65}, 1697-1700 (1990).
\bibitem[]{GeometricMeaning}D. C. Brody, %Elementary Derivation for Passage Times
Journal of Physics A: Mathematical and General \textbf{36}, 5587-5593 (2003).


\bibitem[]{Giovannetti11}V. Giovannetti, S. Lloyd, and L. Maccone, %Quantum Measurement Bounds Beyond the Uncertainty Relations
  arXiv:1109.5661 (2011).


\bibitem[]{pezze09}L. Pezz\'e and A. Smerzi,% Entanglement, Nonlinear Dynamics, and the Heisenberg Limit
Phys. Rev. Lett. \textbf{102}, 100401 (2009).



\bibitem[]{EntanglementFisher}P. Hyllus, W. Laskowski, R. Krischek, C. Schwemmer, W. Wieczorek, H. Weinfurter, L. Pezz\'e, and A. Smerzi, %Fisher Information and Multiparticle Entanglement
Phys. Rev. A \textbf{85}, 022321 (2012);
G. T\'oth, %$Multipartite Entanglement and High-precision Metrology Phys. Rev. A 85,
\textit{ibid.}, 022322 (2012).


\bibitem[]{Uhlmann}A. Uhlmann, %An Energy Dispersion Estimate
Physics Letters A \textbf{161}, 329-331 (1992).


\bibitem[]{BraunsteinCaves96}S. L. Braunstein, C. M. Caves, and G. J. Milburn, %Generalized Uncertainty Relations: Theory, Examples, and Lorentz Invariance
Ann Phys (Amsterdam, Neth.) \textbf{247}, 135-173 (1996).


\bibitem[]{Giovannetti03a}V. Giovannetti, S. Lloyd, L. Maccone, %\textit{Fluctuations and Noise in Photonics and Quantum Optics}, proceedings of the SPIE, Volume \textbf{5111}, pp. 1-6 (2003); %V. Giovannetti, S. Lloyd, and L. Maccone, Quantum Limits to Dynamical Evolution
Phys. Rev. A \textbf{67}, 052109 (2003).

\bibitem[]{FrGrPe08}F. Fr\"owis, G. Gr\"ubl, and M. Penz, %Fleming's Bound for the Decay of Mixed States
Journal of Physics A: Mathematical and Theoretical \textbf{41}, 405201 (2008).


\bibitem[]{TraceNorm}The definition of the trace norm for a linear operator $M$ reads $\lVert M \rVert_1 \mathrel{\mathop:}= \mathrm{Tr}\sqrt{M M^{\dag}}$.

  
\bibitem[]{NielsenChuang}M. A. Nielsen and I. L. Chuang, \textit{Quantum Computation and Quantum Information} (Cambridge University Press, Cambridge, 2010).

\bibitem[]{Bures}D. Bures, %An Extension of Kakutani's Theorem on Infinite Product Measures to the Tensor Product of Semifinite w*-Algebras
T. Am. Math. Soc. \textbf{135}, 199-212 (1969).

 \bibitem[]{Uhlmann86}A. Uhlmann, %Parallel Transport and ``quantum Holonomy'' Along Density Operators
Rep. Math. Phys. \textbf{24}, 229-240 (1986).

 

\bibitem[]{Heisenberg}For two self-adjoint operators $A$ and $B$ and the density operator $\rho$, the Heisenberg uncertainty relation (in the form proved by Robertson) reads $(\Delta A)_{\rho}(\Delta B)_{\rho} \geq \frac{1}{2}\left| \langle \left[ A,B \right] \rangle_{\rho} \right|$. Here, $A=H$ and $B=\Pi_{\rho}$.

 
\end{thebibliography}
\end{document}